\documentclass[twocolumn,showpacs,preprintnumbers]{revtex4}%
\usepackage{graphicx}
\usepackage{dcolumn}
\usepackage{bm}
\usepackage{amsmath}
\usepackage{amsfonts}
\usepackage{amssymb}%

\providecommand{\U}[1]{\protect\rule{.1in}{.1in}}
\begin{document}
\title{Comment on \textquotedblleft Existence of Einstein static universes and their
stability in fourth-order theories of gravity\textquotedblright}
\author{John Miritzis}

 \affiliation{Department of Marine Sciences, University of the
Aegean\\University Hill, Mytilene 81100, Greece\\
 E-mail: imyr@aegean.gr}
\date{\today}

\begin{abstract}
It is argued that the solution space of the equation determining the class of $f(R)$
theories which admit an Einstein static universe should be broadened by including the
algebraic roots.

\end{abstract}

 \maketitle

In a recent paper, Goswami, Goheer and Dunsby (GGD) study the existence of Einstein
static universes in the context of higher order theories of gravity
\cite{ggd}. They consider theories derived from a Lagrangian density%
\[
L=f\left(  R\right)  -2\Lambda+L_{m}
\]
in the Einstein static universe. The field equations imply Eq. (14) in Ref.
\cite{ggd}, namely,%
\begin{equation}
Rf^{\prime}-\frac{3}{2}\left(  1+w\right)  f+3\Lambda\left(  1+w\right)  =0,
\label{equation}%
\end{equation}
where $w\in\left[  -1,1\right]  $ is the parameter in the equation of state,
$p=w\rho$. Equation (\ref{equation}) has been discovered several times (cf \cite{baot}
for a brief discussion of static solutions). GGD argue that the only functions
$f\left(  R\right)  $ admitting the existence of an Einstein static universe are
solutions of the \emph{differential equation} (\ref{equation}) and therefore, have the
form
\begin{equation}
f\left(  R\right)  =2\Lambda+KR^{\frac{3}{2}\left(  1+w\right)  }+L_{m}.
\label{sodi}%
\end{equation}

According to this analysis, the solution (\ref{sodi}) leaves no room for the theory
derived from the Lagrangian density
\[
L=R+\beta R^{2}+L_{m},
\]
to admit the Einstein static universe as a solution. However, in a previous
investigation it was found that this is not the case \cite{miri}. More precisely,
without making use of expansion-normalized variables it is shown in Ref. \cite{miri}
that the dynamical system has an unstable equilibrium of saddle type corresponding to
the Einstein static universe. What is most interesting is that the effective
cosmological constant is not assumed from the beginning, but is provided by the
curvature equilibrium
\begin{equation}
R_{\ast}=\frac{1+3w}{\beta\left(  1-3w\right)  }, \label{eq1}%
\end{equation}
in the range $-\frac{1}{3}<w<\frac{1}{3}$.

In the conclusions of Ref. \cite{ggd}, GGD consider as surprising the result that only
Lagrangians of the form $f\left(  R\right)  =a+bR^{c}$ admit an Einstein static space
and furthermore comment: \textquotedblleft This result differs from some recent work
[36] (Ref. \cite{bhl} in this comment), where the stability of the Einstein static
solution was investigated for a number of types of $f\left(  R\right)  $ theories
which appear not to fall into the class discovered here\textquotedblright.

The existence of the Einstein static universe in $f\left(  R\right)  $ theories which
do not to fall into the class (\ref{sodi}), relies on the interpretation of
(\ref{equation}). In fact, the solution (\ref{sodi}) to the equation (\ref{equation})
is not the only possibility. Given an arbitrary differentiable function $f$, Eq.
(\ref{equation}) can be regarded as an algebraic equation to be solved for $R$.
Denoting the resulting roots by $\rho_{1},\rho_{2},...$ one obtains a whole series of
Einstein spaces, each having a constant scalar curvature $\rho_{i}$ (see for example
the comments in Ref. \cite{cmq} and for a thorough analysis see Ref. \cite{ffv}). In
view of the above interpretation, it is easy to check that substituting $f\left(
R\right) =R+\beta R^{2}$ in (\ref{equation}) with $\Lambda=0,$ the scalar curvature
takes the constant value given by (\ref{eq1}), exactly as was shown in Ref.
\cite{miri}.

\end{document}